# Low-surface-brightness spheroidal galaxies as Milgromian isothermal spheres

R. H. Sanders★

*Kapteyn Astronomical Institute, P.O. Box 800, NL-9700 AV Groningen, the Netherlands*

**ABSTRACT**
I consider a sample of eight pressure-supported low-surface-brightness galaxies in terms of Milgrom's modified Newtonian dynamics (MOND). These objects include seven nearby dwarf spheroidal galaxies – Sextans, Carina, Leo II, Sculptor, Draco, Leo I, Fornax, and the ultra-diffuse galaxy DF44. The objects are modelled as Milgromian isotropic isothermal spheres characterized by two parameters that are constrained by observations: the constant line-of-sight velocity dispersion and the central surface density. The velocity dispersion determines the total mass, and, with the implied mass-to-light ratio, the central surface brightness. This then specifies the radial run of surface brightness over the entire isothermal sphere. For these objects, the predicted radial distribution of surface brightness is shown to be entirely consistent with observations. This constitutes a success for MOND that is independent of the reduced dynamical mass.

**Key words:** galaxies: dwarf – dark matter.

## 1 INTRODUCTION

In the demographics of galactic scale objects, dwarf spheroidals, along with their gassy cousins the dwarf irregular galaxies, are the most common members of the galaxy population. As such, they are relevant to cosmological issues such as their role in galaxy formation – the hierarchical build-up of more massive objects through successive merging. These objects are abundantly predicted in cold dark matter cosmological *N*-body simulations – in fact, in larger abundance than actually observed – and so they are significant for dark matter scenarios of the evolution of structure. Primarily, for this reason, these objects have garnered considerable attention among astronomers and cosmologists (see Mateo 1998 for an older but excellent review).

For decades, it has been appreciated that low-surface-brightness (LSB) objects exhibit a large discrepancy between the conventional luminous mass and the total Newtonian dynamical mass with mass-to-light ratios in several case exceeding 100 in solar units. In dark matter cosmogony, this is attributed to the role of stellar feedback – winds, supernovae, etc. – in removing baryonic matter from these comparatively shallow potential wells. Indeed, in this picture, the predicted but unobserved dwarfs are presumed to be present as purely dark objects.

But there is another solution to the dark matter dominance in these objects, and that is due to existent dynamics rather than evolutionary processes. This alternative is Milgrom's modified Newtonian dynamics (MOND) proposed by Milgrom (1983a,b). In his initial papers, Milgrom suggested that in astronomical systems the observed discrepancy between the visible and conventional dynamical mass appears below a critical acceleration, $a_0$, and so proposed MOND as an acceleration-based modification of Newtonian gravity or dynamics. But surface density is a moniker for acceleration, so an immediate prediction of the theory is that the mass discrepancy should be conspicuous in low surface density or, equivalently, LSB systems. This is to say that LSB objects such as dwarf spheroidal galaxies should apparently exhibit a large ratio of dark-to-visible matter (a true prediction because it was stated before there was substantial evidence for the high dark matter content in these objects). That this phenomenology is unrelated to the small size of these objects is supported by the observations of large LSB galaxies such as Malin 1 (Bothun et al. 1990) and the recently discussed ultra-diffuse galaxies (van Dokkum et al. 2019) where a significant mass discrepancy appears in Milky Way-size galaxies. This implicates the surface density, or acceleration, as the causative factor. At the other extreme, there are high-surface-brightness objects such as globular clusters and ultra-compact galaxies that exhibit no discernible discrepancy within the visible image (e.g. Scarpa 2005).

Here, I consider the possibility that the structure of LSB spheroidal galaxies is described by deep-MOND (low internal accelerations) self-gravitating spheres by making three greatly simplifying and constraining assumptions: perfect spherical symmetry, complete isotropy of the velocity field, and a constant velocity dispersion (isothermality). Further, I do not consider complications such as tides, the MOND external field effect (EFE) (discussed in final section), or rotation of the system (although see Martínez-García et al. 2021). With these assumptions, the structure of the system is completely determined by two observed quantities: the line-of-sight velocity dispersion and the central surface brightness. Then for a sample of eight LSB galaxies, I demonstrate that the observed radial run of surface density is matched by that of the corresponding Milgromian isothermal sphere.

★ E-mail: sanders@astro.rug.nl

## 2 STRUCTURE OF MILGROMIAN ISOTHERMAL SPHERES

In the original (primitive) MOND, the form of the gravitational force on a particle ($g$) is given by

$$\mu(g/a_0)g = g_N \qquad (1)$$

and $g_N$ is the usual Newtonian acceleration related to the density distribution via the usual Poisson equation. The interpolating function $\mu$ is not specified but must have specific asymptotic forms: $\mu(x) = 1$ when $x \gg 1$ (the Newtonian limit) and $\mu(x) = x$ when $x \ll 1$ (the Milgromian limit). With spherical symmetry, this simple expression is also valid for physically viable formulations of MOND as a modification of gravity: i.e. the aquadratic Lagrangian theory of Bekenstein & Milgrom (1984) and the quasi-linear theory (QMOND) of Milgrom (2010).

In the low acceleration regime, the effective force becomes $g = \sqrt{g_N a_0}$. Sufficiently distant from a finite object this may be written as $g = \sqrt{GMa_0}/r$ leading to asymptotically flat rotation curves and a rotation velocity–mass relationship (Tully–Fisher) of the form $V^4 = GMa_0$ as observed in spiral galaxies. Apart from these general relations, the MOND simple formula has been shown to precisely account for the rotation curves of spiral galaxies from the observed distribution of baryonic matter in a large number of cases (Li, Lelli, McGaugh, Schombert 2018).

The structure of self-gravitating pressure-supported objects such as elliptical galaxies and dwarf spheroidals in the context of MOND was first addressed by Milgrom (1984) in a discussion of isothermal spheres. A well-known historical result is that with Newtonian dynamics (Chandrasekhar 1957) a self-gravitating isothermal gas sphere has infinite extent (density falling as $1/r^2$) and infinite mass. But Milgrom demonstrated that with MOND such an object has finite mass (asymptotic density falling as roughly $1/r^4$) essentially because of the effectively stronger gravity at low acceleration. This mass is proportional $\sigma_l^4/Ga_0$ where $\sigma_l$ is the line-of-sight velocity dispersion averaged over the entire system. This forms the basis of the observed velocity dispersion–luminosity relationship (Faber–Jackson) $L \propto \sigma^4$ (see Sanders 2010 for a discussion of the applicability of the Faber–Jackson relation from dwarf spheroidals to clusters of galaxies).

The structure equation for the isothermal sphere is

$$\sigma_r^2 \left(\frac{d\ln(\rho)}{dr} + 2\frac{\beta}{r}\right) = -g, \qquad (2)$$

where $g$ is the modified gravitational force given by equation (1) and the anisotropy parameter is $\beta = 1 - \sigma_t^2/\sigma_r^2$ where $\sigma_r$ and $\sigma_t$ are the radial and tangential velocity dispersion, respectively; $\beta$ may range from $-\infty$ (pure circular orbits) to $+1$ (radial orbits), but here I will consider only isotropic systems ($\beta = 0$).

It is convenient to rewrite the equation in dimensionless form using scale parameters for mass, radius, density, and surface density:

$$\tilde{M} = \sigma_r^4 (a_0 G)^{-1}, \quad \tilde{r} = \sigma_r^2 a_0^{-1},$$
$$\tilde{\rho} = a_0^2 (4\pi G \sigma_r^2)^{-1}, \quad \tilde{\Sigma} = a_0/G. \qquad (3)$$

Milgrom demonstrated that there is a family of solutions (scaled by $\sigma_r$) and the parameter $\alpha$ that is the central density in terms of the density scaling ($\rho_0/\tilde{\rho}$ above) ranging from infinity (a limiting or cusp-like solution) and solutions with finite $\alpha$ having a constant density core.

Luminous elliptical galaxies are not well represented by Milgromian isotropic isothermal spheres (Sanders 2000). The line-of-sight velocity dispersions generally decline with radius and the mass discrepancy is often vanishingly small within a few effective radii, i.e. accelerations are greater than $a_0$ in these inner regions (Milgrom & Sanders 2003; Romanowsky et al. 2003). To model such systems in the context of MOND, it is necessary to exploit other possible degrees of freedom such as the anisotropy parameter (possibly variable) and deviations from isothermality expressed, for example, by a polytropic pressure–density relation, $\sigma_r^2 \propto \rho^{1/n_p}$, where $n_p$ is the polytropic index. A reasonable representation of the observed properties of these systems, including the Fundamental Plane (the observed virial relation between mass, velocity dispersion, and effective radius), can be described by anisotropic ($\beta > 0$), near-isothermal polytropes with $n_p \approx 12-16$ (Sanders 2000).

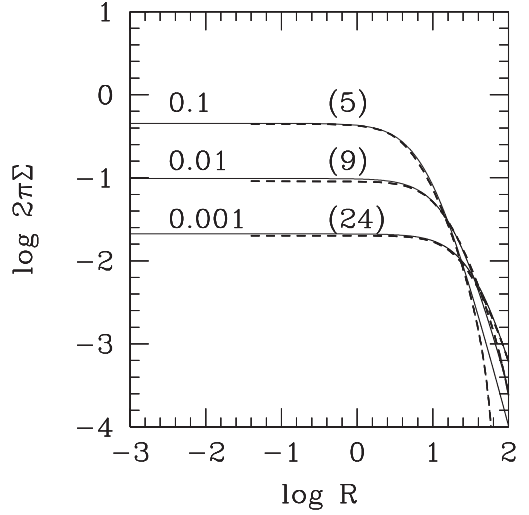

**Figure 1.** The surface density of Milgromian isothermal isotropic spheres compared to King models where the units of radius and surface density are in terms of the MOND units (equation 3). The various solid curves correspond to the indicated values of $\alpha = \rho_0/\tilde{\rho}$, the only free parameter for isotropic isothermal models. The dashed curves are the best corresponding empirical King models (King 1962) where the core radius is shown in parenthesis and the cut-off radius is taken to be 14 times the core radius.

The cored isothermal spheres with $\alpha < 1$ are generally deep MOND objects with internal acceleration everywhere less than $a_0$, and as such one might expect that these simpler systems may provide a reasonable description of LSB spheroidal systems. Restricting to strictly isotropic systems ($\beta = 0$) then asymptotically (as $r \to \infty$ and $M \to$ constant) equation (2) becomes

$$\frac{d\ln(\rho)}{d\ln(r)} = -\sqrt{GMa_0/\sigma_r^4} = -Q. \qquad (4)$$

For such deep MOND systems, Milgrom found that $Q = 9/2$; that is to say, the asymptotic density distribution is a power law with slope $-9/2$. This implies

$$M = \frac{81}{4} \sigma_l^4 (Ga_0)^{-1} \qquad (5)$$

given that $\sigma_l = \sigma_r$ in isotropic systems (Milgrom 1984).

To allow for possible deviations from deep MOND, I have here recalculated the structure of pressure-supported Milgromian spheres by numerically solving the structure equation with $\mu(x) = x/(1 + x^2)^{0.5}$. The distribution of surface density in the cored solutions resembles that of the empirical King models (King 1962) that have been used to fit the observed run of surface brightness in globular clusters and dwarf spheroidals. In Fig. 1, the solid curves show the run of scaled surface density for the isothermal spheres as a function of scaled radius for three different members of the family characterized by $\alpha$. The dashed curves show the best corresponding King models characterized by two parameters – a core length-scale and a cut-off radius that is taken here to be 14 core radii. There is a close agreement between the MOND isothermal spheres and the King models given that isothermal spheres continue as a power law ($-4.5$) to infinity while the King models truncate at the 'tidal' radius. This demonstrates already that, in so far as King models fit the surface brightness distribution of LSB spheroidal systems, so will Milgromian isothermal spheres.

**Table 1.** LSB galaxies.

| Galaxy | $L_V$ ($10^6$ L$_\odot$) | $\sigma_1$ (km s$^{-1}$) | SB$_0$ (mag arcsec$^{-2}$) | $\alpha$ $\rho_0/\tilde{\rho}$ | $M_s$ ($10^6$ M$_\odot$) | $r_{\rm eff}$ (kpc) | $\Sigma_0/\tilde{\Sigma}$ | (M/L)$_V$ |
|---|---|---|---|---|---|---|---|---|
| Sextans | 0.44 | 6.1 | 27.3 | 0.00045 | 1.75 | 0.569 | 0.0176 | 4.0 |
| Carina | 0.38 | 6.6 | 25.4 | 0.012 | 2.4 | 0.369 | 0.0197 | 6.3 |
| Leo II | 0.74 | 6.6 | 24.0 | 0.035 | 2.40 | 0.246 | 0.036 | 3.24 |
| Sculpter | 2.3 | 8.1 | 23.7 | 0.03 | 5.45 | 0.339 | 0.032 | 2.4 |
| Draco | 0.29 | 8.5 | 25.0 | 0.13 | 6.61 | 0.209 | 0.084 | 22.8 |
| Leo I | 5.5 | 9.7 | 22.4 | 0.1 | 11.2 | 0.290 | 0.0718 | 2.04 |
| Fornax | 20 | 9.8 | 23.3 | 0.005 | 12.7 | 0.756 | 0.012 | 0.8 |
| DF 44 | 300 | 28 | 24.4 | 0.015 | 778 | 6.568 | 0.0294 | 2.6 |

To apply the model to actual objects, I make use of the average measured line-of-sight velocity dispersion in equation (5) to estimate the total mass and mass-to-light ratio of the corresponding deep MOND isotropic isothermal sphere. Then, there is only one adjustable parameter: $\alpha = \rho_0/\tilde{\rho}$, the cored density in terms of the MOND density. So the problem for a particular object is to set $\alpha$ such that, with the implied mass-to-light ratio, the central surface brightness matches that observed. The predicted run surface brightness is then that of the isothermal sphere.

## 3 THE SURFACE BRIGHTNESS PROFILES OF LSB GALAXIES

The sample consists of eight pressure-supported LSB objects: seven are well-observed LSB dwarf spheroidal satellites of the Milky Way and one is a recently discussed larger ultra-diffuse galaxy at a distance of 100 Mpc. These are chosen because the surface brightness profile has been observed with reasonable spatial resolution (Irwin & Hatzidimitriou 1995; van Dokkum et al. 2015). All objects are modelled as spherically symmetric MOND isothermal spheres with isotropic velocity distributions. I do not consider the MOND EFE (this is discussed in the final section). The sample of objects is given in Table 1 along with observed and derived parameters.

In Fig. 2, I show for the eight objects the observed profile of line-of-sight velocity dispersion along with adapted $\sigma_1$ for each object given in order of increasing $\sigma_1$. Because the objects are assumed to be isothermal and isotropic, the adapted $\sigma_1$ appears as a horizontal line; in most cases, this provides a reasonable approximation to the observations.

In Table 1, the line-of-sight velocity dispersion (column 3) ranges between 6 km s$^{-1}$ (Sextans dwarf spheroidal) and 30 km s$^{-1}$ (Dragonfly 44). The central surface brightness (column 4) ranges from 27.3 (Sextans) to 22.4 (Leo I) in units of V magnitude per square arc second. In fitting the surface brightness profiles, I only adjust $\alpha$, the central core density with respect to the MOND density. This is the parameter that determines the form of the surface density distribution (the observed velocity dispersion determines the amplitude of the distribution). The isothermal sphere mass, effective radius, the central surface density in units of $a_0/G$, and the implied mass-to-light ratio are given for the corresponding MOND model. In these units, the central surface densities range from 0.018 (Sextans) to 0.07 (Leo I), so all objects are completely in the deep MOND regime. The mass-to-light ratios are mostly consistent with normal stellar populations except for Draco (23.4).

For all objects, the model surface brightness profiles (solid curves) are compared to observations (points) in Fig. 3. These models are scaled from the calculated surface *density* distributions for the appropriate isothermal sphere (specified by $\sigma_1$ and the central density $\rho_0$) using the implied global mass-to-light ratios. The relationship between the surface brightness distribution, SB($r$) in V magnitudes per square arcsecond, is then given by

$$\text{SB}(r) = 26.4 - 2.5\log\left[\frac{\Sigma(r)}{M/L_V}\right], \quad (6)$$

where $\Sigma(r)$ is the radial run of surface density for the appropriate isothermal sphere (in units of M$_\odot$ pc$^{-2}$ and M/L$_V$ is the estimated M/L for the particular object (column 9, Table 1). The observed run of surface brightness is taken from the observations of the surface number density of luminous stars (Irwin & Hatzidimitriou 1995) converted to magnitudes per arcsecond and normalized to the central surface brightness (column 4, Table 1).

We see that the surface brightness profiles for the isothermal spheres provide a reasonable match to the observations particularly considering that there is only one adjustable free parameter, $\alpha$, that is constrained by the observed central surface brightness. That is to say, by adjusting $\alpha$ in order to match the observed surface brightness at one point, the centre, the entire radial run of surface brightness for the object is specified by the corresponding Milgromian isothermal sphere and found to agree with the observations to some precision.

In the case of the Fornax dwarf, the observations are not well matched by surface density distribution of the best-fitting isothermal sphere. Moreover, the observed radial decrease of the line-of-sight velocity dispersion is also not consistent with an isotropic isothermal sphere. In this case, I make use of the additional degree of freedom of a pressure–density relation, the polytropic relation, where I take the polytropic index to be $n_p = 8$. In Fig. 4, we see that the run of velocity dispersion and surface brightness is well matched and the overall implied M/L remains 0.8. (A high-order polytrope, $n_p \approx 12$, also provides a better fit to the velocity dispersion and surface brightness profiles of Leo II than does an isothermal sphere.)

With respect to the ultra-diffuse galaxy Df44, Bílek et al. (2019) have recently carried out Jeans modelling of this object but followed a different procedure. They take a Sersic model fitted to the light distribution and assigned an M/L of 1.9 to the galaxy (a mass of $3.9 \times 10^8$ M$_\odot$). From the structure equation, they derive a velocity dispersion and found that it near isothermal at approximately 24 km s$^{-1}$. This is consistent with the result given here. I assumed a velocity dispersion of 28 km s$^{-1}$ as a reasonable match to the line-of-sight velocity distribution; the corresponding Milgromian isothermal fit (with $\alpha = 0.015$) then matches the central surface brightness and the observed surface brightness distribution over the entire system. The higher mass of $7.2 \times 10^8$ M$_\odot$ is consistent with the MOND $M \propto \sigma^4$ relation and the implied M/L, 2.5, is reasonable for a stellar population (see also Haghi et al. for a recent discussion of this object from the point of view of MOND).

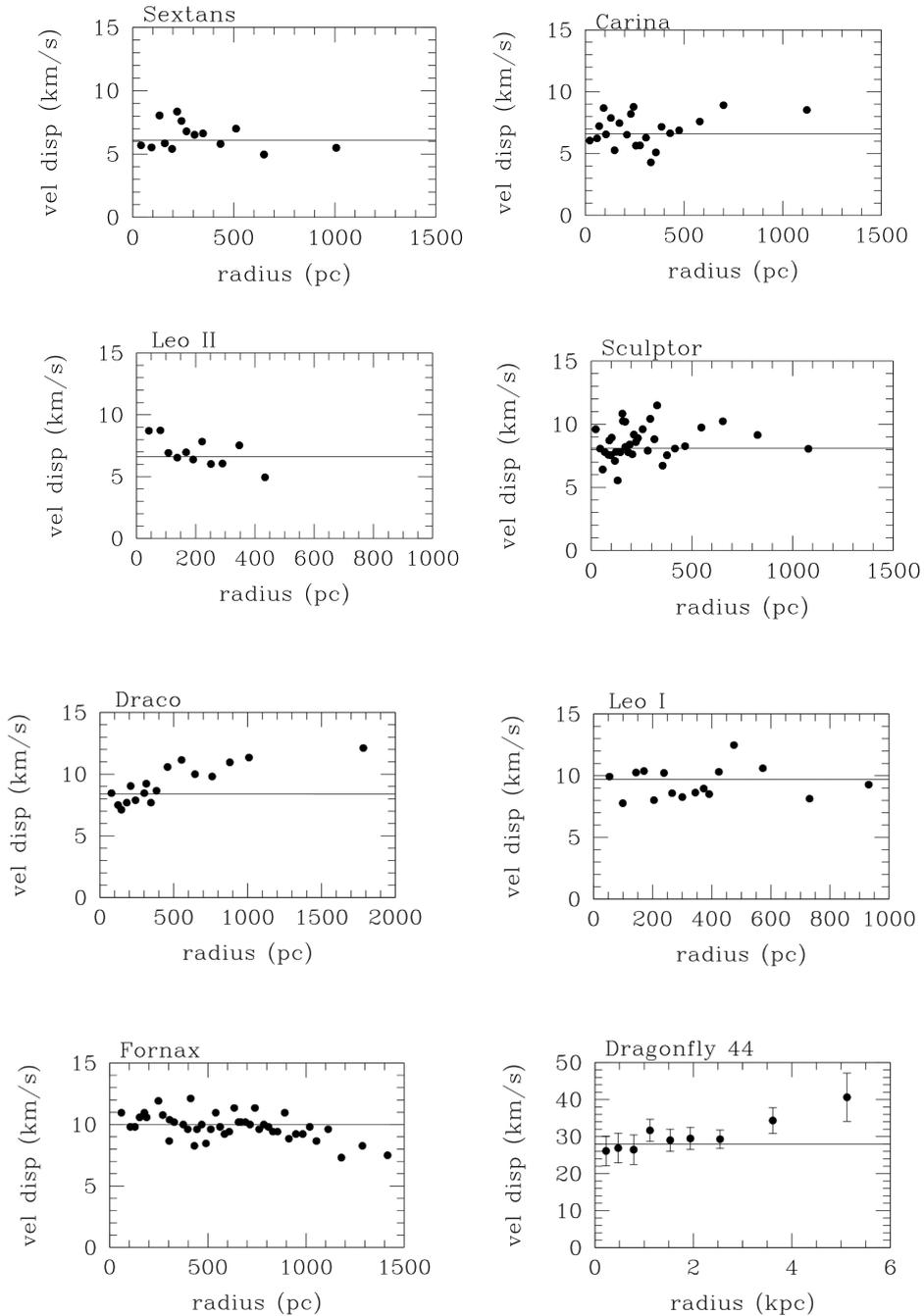

**Figure 2.** Observed velocity dispersion profiles of the sample galaxies and the line-of-sight velocity dispersion for the adapted isothermal spheres. Dwarf spheroidal data are from Walker et al. (2009); DF44 data are from van Dokkum et al. (2019). The horizontal lines are the line-of-sight velocity dispersion of the corresponding isothermal spheres.

## 4 DISCUSSION

Most of the discussion of dwarf spheroidal galaxies in the context of MOND has concerned several estimated high mass-to-light ratios even with MOND. For example, Gerhard & Spergel (1992), basically with equation (5) applied to the sample of Milky Way dwarfs with velocity dispersion data as observed at that time, suggested a wide range of values for the mass-to-light ratios implied by MOND – up to 30 in solar units – and they argued that this could be taken as evidence against MOND. Milgrom (1995) responded that the errors, primarily on the measured velocity dispersion, were sufficiently large that sensible mass-to-light ratios were also possible, and indeed later observations of lower measured velocity dispersion have generally decreased the estimated M/L values.

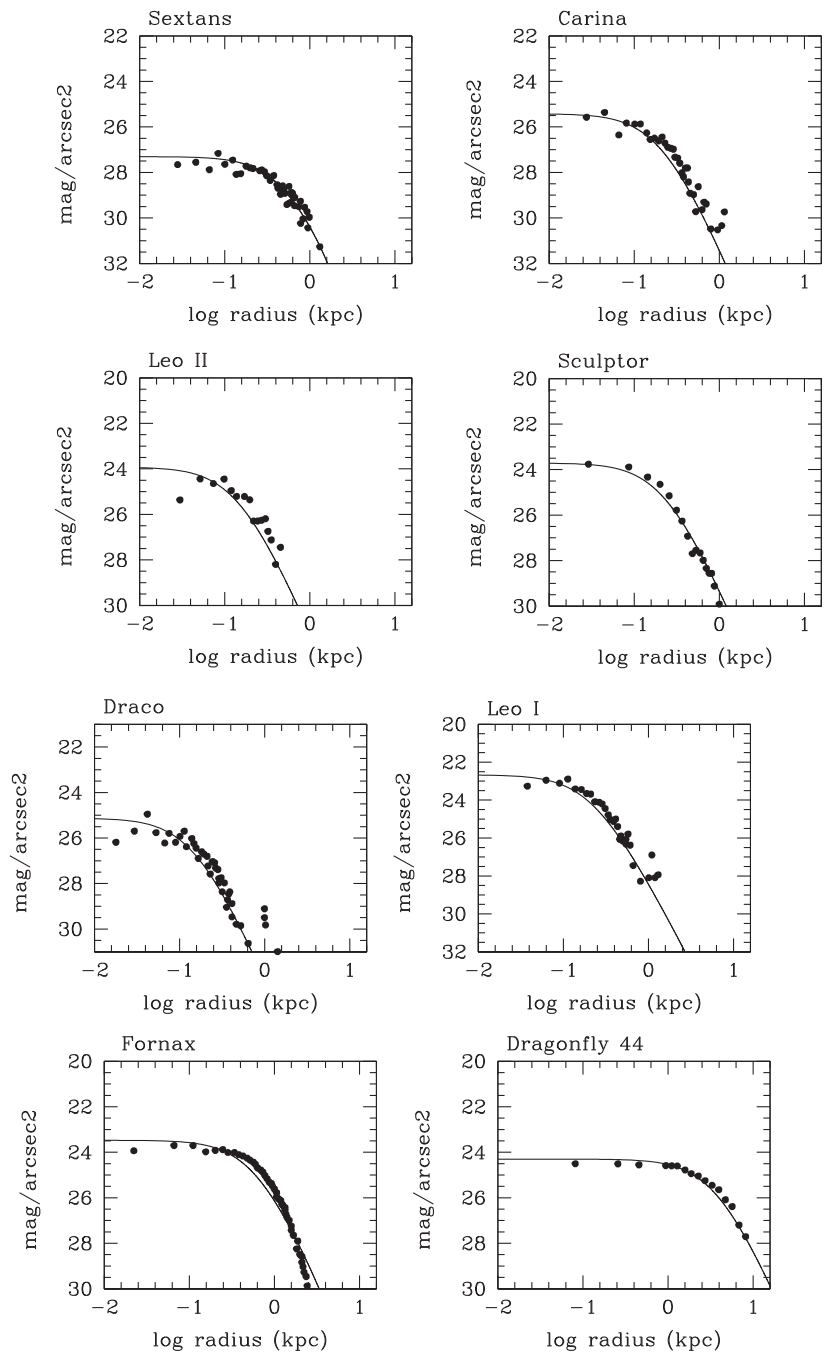

**Figure 3.** Surface brightness profiles for sample galaxies. The points are the data (Irwin & Hatzidimitriou 1995 for dwarf spheroidals and van Dokkum et al. 2019 for Df 44). The curves are the isothermal sphere models with central surface brightness set from the estimated M/L ratios.

By fitting the radial dependence of $\sigma_1$, Angus (2008) stressed the possible role of a variable anisotropy parameter and concluded that only in two cases is implied M/L sufficiently higher (>10) than normally expected for stellar populations (in particular Draco with M/L ≈ 45). In these cases, Angus considered possible solutions – non-equilibrium, impulsive heating, tidal disruption, tidal tails and ellipticity, and dust. Recent discussions of the effect of the stellar population on the elevated Milgromian M/L of Carina is given by Angus et al. (2014), and of the general effect of tides on Draco among other objects by McGaugh & Wolf (2010).

Brada & Milgrom (2000), using numerical solutions to the Bekenstein–Milgrom field equation (Bekenstein & Milgrom 1984) (considering MOND as a modification of Newtonian gravity) made the first detailed study of the EFE. The EFE is an aspect MOND as a non-linear theory, and implies that an external gravitational field can affect the internal dynamics of a system independently of tides even if the external acceleration field is constant. Basically, when the external field dominates, then in the Milgromian regime ($a_0 > g_{\text{ext}} > g_{\text{int}}$) the dynamics is Newtonian but with a larger effective constant of gravity (by a factor of $a_0/g_{ext}$). Brada & Milgrom demonstrated that dwarf satellite galaxies, being on elongated orbits, may explore several distinct regimes: external/internal field dominated, adiabatic or impulsive response of the satellite, or tidal field dominated and this of course can complicate the interpretation of the observations. This issue has been recently revisited in numerical calculations by Lüghausen, Famaey & Kroupa (2014). They found that the implied M/L values for Sextans, Carina, and Draco remain problematic for MOND but considered several escape clauses. One of these is that MOND may be a modification of inertia rather than gravity. Milgrom (1994, 2011) has demonstrated that any action-based theory of MOND as modified inertia must be non-local in time implying that the internal dynamics depends upon the entire trajectory of the system through phase space and not at only one point on that trajectory. This could have a dramatic effect on the presence of the EFE particularly for systems moving on highly elongated orbits as are the dwarf spheroidals (see Chae et al. 2020 for a discussion of the presence of the EFE as revealed by precise galaxy rotation curves).

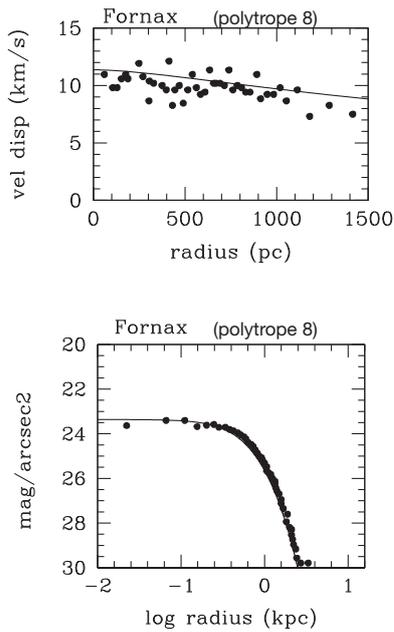

**Figure 4.** Polytropic ($n_p = 8$) fit to Fornax velocity dispersion and surface brightness profiles.

Here, I have ignored the EFE and other complications such as deviations from spherical symmetry, isothermality, and isotropy. The issue is simply to determine how well the MOND isotropic isothermal sphere reproduces the run of surface brightness in LSB stellar systems that are deep within the low-acceleration regime. In a sense, it is similar to the problem of matching rotation curves of spiral galaxies – essentially a one-parameter fit to a measured one-dimensional function of radius. Indeed the fitted parameter ($\alpha$) that for a given $\sigma_1$ specifies the central surface density is constrained by matching the measured central surface brightness using the M/L ratio implied by the velocity dispersion.

We see (Fig. 3) that this procedure works quite well in describing the run of surface brightness in terms of this highly constrained model. In one case where there are systematic deviations between the model and observations, (Fornax) a minimal departure from isothermality, a high-order polytrope, can produce a near perfect match to both the radial run of velocity dispersion and surface brightness distribution (Fig. 4). Of course, this procedure is not a powerful as the matching of rotation curves; the rotation curves provide a more direct and precise measure of the run of radial acceleration in a galaxy as expressed by the 'radial acceleration relation' (McGaugh et al. 2016). But overall this result is additional evidence for the relevance of modified dynamics to these near perfect Milgromian systems.

## ACKNOWLEDGEMENTS

I am grateful to Moti Milgrom for many helpful comments on the original manuscript. I also thank M. Walker, P. van Dokkum, and M. Irwin for permission to use data published in graphic form.

## DATA AVAILABILITY

Data sets for this research are included in Walker et al. (2009, fig. 1), Irwin & Hatzidimitriou (1995, fig. 2), Mateo (1998, table 3), McConnachie (2012, table 3), and van Dokkum et al. (2015, fig. 5, 2019, fig. 14).

This paper has been typeset from a T<sub>E</sub>X/L<sup>A</sup>T<sub>E</sub>X file prepared by the author.